# Irradiation of Nuclear Emulsions in Relativistic Beams of $^6$He and $^3$H Nuclei

M. I. Adamovich$^a$, A. M. Baldin$^{a,b}$, V. Bradnova$^b$, M. M. Chernyavsky$^a$, V. A. Dronov$^a$,
S. P. Kharlamov$^a$, A. D. Kovalenko$^b$, V. A. Krasnov$^a$, V. G. Larionova$^a$, G. I. Orlova$^a$,
N. G. Peresadko$^a$, P. A. Rukoyatkin$^b$, V. V. Rusakova$^b$, N. A. Salmanova$^a$, P. I. Zarubin$^b$

$^a$ P. N. Lebedev Physical Institute (FIAN), Russian Academy of Sciences, Moscow
$^b$ Laboratory of High Energies, Joint Institute for Nuclear Research, Dubna

**Abstract**

Using a $^6$Li nucleus beam extracted from the JINR Synchrophasotron at momentum 2.67A GeV/c a secondary beam was produced with a composition of 1% of $^6$He and 99% of $^3$H nuclei. Preliminary results on the features of nucleus-nucleus interactions of $^6$He nuclei and charge exchange (CE) of $^3$H nucleus are presented. Interactions of $^6$He nucleus external neutrons with emulsion nuclei as well as a coherent stripping of $^6$He nucleus external neutrons are observed. A mean range of $^3$H nuclei to inelastic interactions in emulsion is equal to 23.7±3.0 cm.

A mean range in emulsion for a nuclear CE process $^3$H→$^3$He is equal to 40±16 m. The CE cross-sections with a charged meson production or without it are approximately equal. The CE cross-section with excitation of a target nucleus exceeds the one without excitation. The mean transverse momentum of $^3$He nuclei is equal to 0.16±0.03 GeV/c.

**Introduction**

Beams of light radioactive nuclei are under development now at the Nuclotron accelerator complex. Nuclei situated near proton and neutron drip lines often have no uniform structure in which separate nucleons or a few nucleon clusters are weakly bound with respect to a dominant nucleus part. Being compared with uniform density nuclei these isotopes have distinctive interaction features. Fragmentation studies of relativistic nuclei having an exotic structure can provide additional information about the structure features.

Application of nuclear emulsions is most efficient for neutron deficit nuclei containing weakly bound protons or charged nucleon clusters. Among the known ones the emulsion technique provides the best spatial resolution. It is possible to register an interaction vertex and all secondary charged particles in emulsion. Thus, an event topology can be defined. Studying fragments of relativistic nuclei produced in various topology interactions one is able to obtain information on momentum and spatial distributions of nucleons and nucleon associations in a nucleus.

In this paper, we have suggested development of a $^6$He nucleus beam and have described a test irradiation with such a beam. The neutron rich $^6$He nucleus is the simplest one having a neutron halo. It has a core presented by α-particle and two external neutrons weakly bound with the core.

The composition and spatial structure of a secondary beam are analyzed and compared with the results obtained by electronic technique in the framework of this study. Besides, topological structures of $^6$He nucleus interactions have been investigated. Such data and, in particular, data on interactions of external nucleons with emulsion is of interest for the development of further experiments on $^6$He nucleus structure. Measurements of multiple Coulomb scattering of $^6$He nucleus tracks provide a direct reference data on the identification of this isotope. Such an analysis has a special significance for studies of CE process $^6$Li→$^6$He, $^6$He nucleus production in fragmentation of $^7$Li and $^7$Be, and dissociation $^{12}$Be→$^6$He$^6$He.

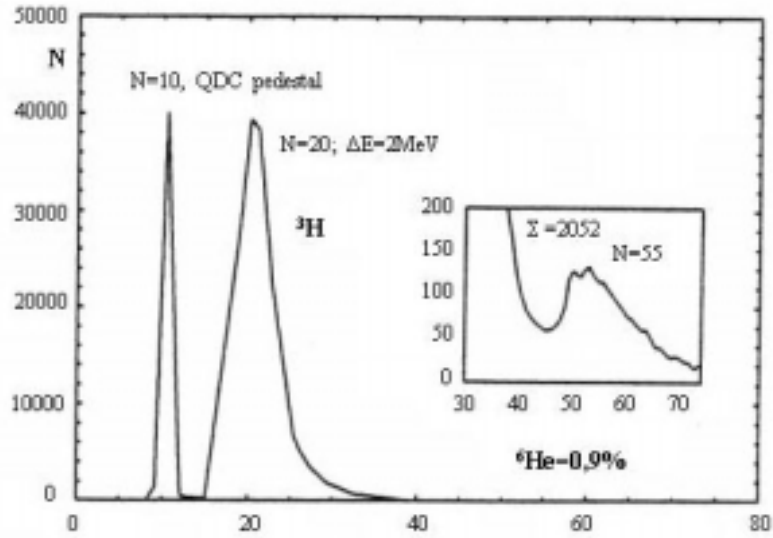

Fig.1 Signal spectrum of one of monitoring scintillation counters for magnetic transport channel tune corresponding to Z/A=1/3

**Secondary beam formation and emulsion irradiation**

Formation of a $^6$He nucleus beam was suggested via a CE process of an extracted $^6$Li nucleus beam. In papers [1-4] the CE process $^6$Li→$^6$He in an emulsion irradiated by a 4.5A GeV/c $^6$Li nucleus beam was observed with a cross-section equal to 4±2 mb.

In this paper $^6$Li nuclei were accelerated in the Synchrophasotron up to pc/Z=5.34 GeV. An extracted $^6$Li nucleus beam was guided to a focal point of a magnetic transport channel. A 4 cm thick plastic target was placed in this point. Secondary particles were selected at an angle 0º by a magnetic channel tuned for Z/A=1/3. In fig. 1 a signal spectrum is shown for one of monitoring scintillation counters. A dominant peak corresponds to $^3$H nuclei. A small peak in large energy losses corresponds to $^6$He nuclei. Fraction of $^6$He nuclei is about 1%. $^6$He nuclei were produced by a CE process of $^6$Li nuclei on target ones (1)

$$^6Li+A \rightarrow {}^6He+X \quad (1).$$

Using the formed beam an emulsion chamber has been irradiated. The chamber was installed in the focal point F6 of the 4B channel. The angular divergence was within an angle 3 mrad. The beam profile for emulsion irradiations was formed in such a way that its horizontal size would correspond to irradiated emulsion width and the spatial beam density would be uniform. The vertical beam size was 3 cm. An emulsion irradiation dose was limited to $10^4$ beam tracks.

The emulsion chamber was assembled as a stack of BR-2 type emulsion layers having sensitivity up to relativistic particles. Layers of thickness 550 µm had the dimensions 10×10 cm$^2$. During irradiation, the beam was directed in parallels to the emulsion plane. Search for double charge particles began at distance 2 cm from enter plane. The found tracks were followed "back" to entrance points and "forward" to interaction vertices. 180 tracks of double charged particles have been found in the scanned layers. Thus, the ratio of the single to double charged particles defined by track counting is equal to 101±14.

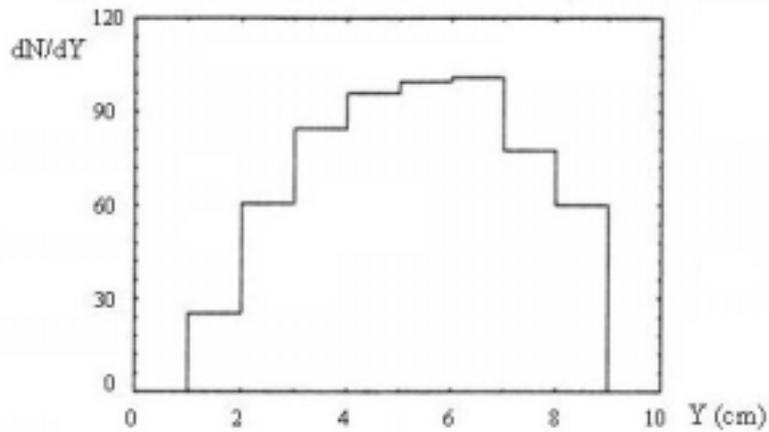

Fig. 2. Horizontal distribution of beam tracks.

The track horizontal distribution (Y coordinate) is presented in fig. 2 for beam single charged particles. The beam tracks are distributed uniformly enough. Due to such a profile of the formed beam the emulsion working area increased by a factor of 3-4 compared with our previous irradiations by smaller size beams.

Angular distributions are presented in the fig. 3 for beam single charged particles, namely, over the angle $\varphi$ in the horizontal projection and over $\alpha$ in the vertical one. A Gaussian fit dispersion is equal to 3.4 mrad for the $\varphi$ distribution, and 2.5 mrad for the $\alpha$ one.

Multiple scattering measurements have been performed and the particle momenta (p$\beta$c) have been defined for 59 double charged beam particles found in five emulsion layers. The obtained p$\beta$c distribution is presented in fig. 4. The measured p$\beta$c values are between 10 and 20 GeV. By Gauss fitting one obtains a mean value $<p\beta c>=15.8\pm0.1$ GeV with $\chi^2 \approx 1$. This value is in good agreement with the beam line momentum. Therefore, double charged beam particles entered though irradiated emulsion side are $^6$He nuclei. Besides, this result confirms that $^6$He nuclei with a momentum of 3 GeV/c is identified reliably enough by the used approach and in such an emulsion.

Summarizing, the emulsion technique confirms the calculated values and measured by monitoring detectors.

### Interactions of $^6$He nuclei with emulsion nuclei

Preliminary results on emulsion visual scanning and measurements of $^6$He nucleus interactions with emulsion nuclei are presented in this section. All found double charged tracks were followed to a nuclear interaction point or to an exit from the emulsion layer. One has found 59 inelastic interactions containing secondary particles over the scanned length of 1114.2 cm of initial double charged particles.

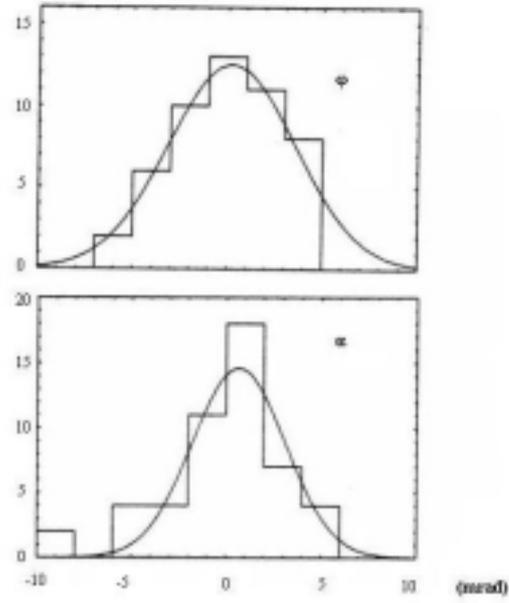

Fig. 3. Angular distributions of beam tracks.

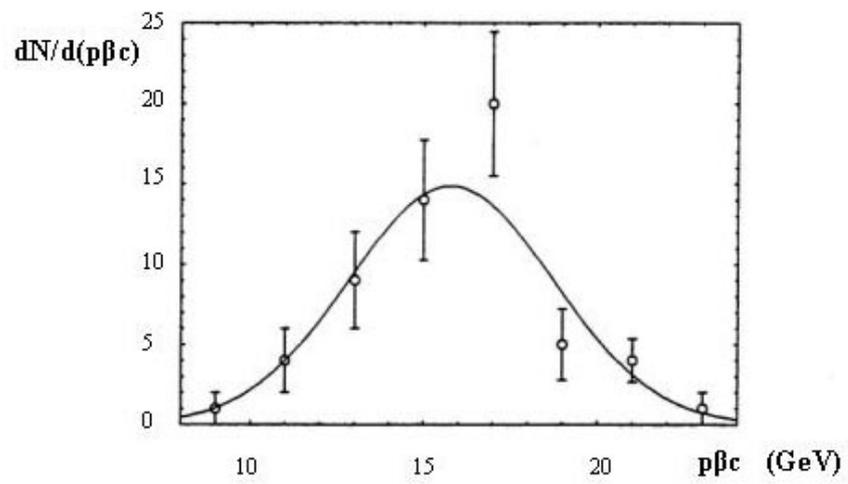

Fig. 4. Reconstructed distribution over pβc for beam double charged tracks.

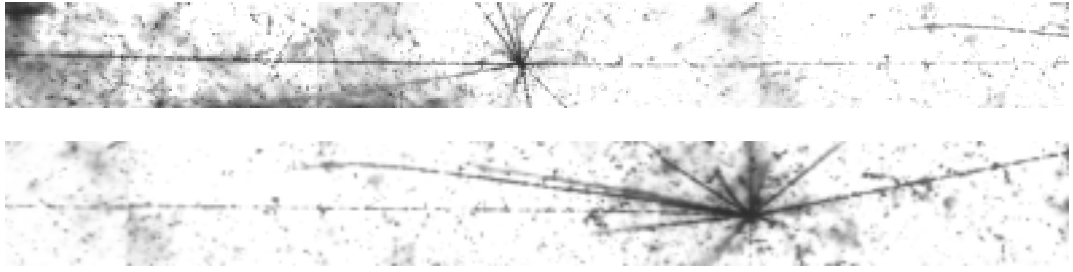

Fig. 5. Event of $^6$He nucleus interaction with fragmentation into α particle. The α particle track is followed until inelastic interaction.

We describe a topological structure of such events. Projectile nucleus fragments are absent in 27 of them. Such fragments are defined as particles with an emission angle θ<6° and having pβc value larger than 1.5 GeV. Mean multiplicity values of charged mesons $<n_\pi>$ and heavily ionizing particles $<n_h>$ are presented for this event sample in table 1.

Table1. Mean multiplicity values in events without projectile nucleus fragments

| Nucleus | $<n_\pi>$ | $<n_h>$ | Momentum (AGeV/c) |
|---|---|---|---|
| $^4$He | 5.6±0.1 | 14.0±0.5 | 4.5 |
| $^6$Li | 10.2±1.0 | 20.1±2.0 | 4.5 |
| $^6$He | 3.9±1.1 | 17.8±3.4 | 2.67 |

The number of mesons is defined as a difference between the total number of relativistic single charged particles with emission angles outside a forward fragmentation cone and the number of projectile nucleus protons participated in an inelastic interaction. The number of $h$ particles includes fragments and cascade particles of a target nucleus. The table includes data on central interactions of $^4$He and $^6$Li nuclei at a primary momentum 4.5A GeV/c.

The mean multiplicity value of charged mesons $<n_\pi>$ produced by $^6$He nuclei is significantly less than the one for $^6$Li interactions and even somewhat less than for the $^4$He case. Mostly this is due to a significant difference in the collision energy value. The value $<n_h>$ reflects a target nucleus fragmentation and has a weak energy dependence in a GeV scale. Its magnitude for the $^6$He nucleus case is in reasonable correspondence with the other ones. The remaining 32 events have projectile nucleus fragments. Table 2 gives the mean multiplicity values for this event sample as well as for $^4$He and $^6$Li nucleus interactions having charged relativistic fragments. A mean value of a total charge of relativistic fragments Q usually serves for estimation of a number of interacted nucleons of a projectile nucleus ν=(Z-Q)A/Z, where A and Z are mass number and charge of a projectile nucleus.

A major feature of $^6$He nucleus interactions with nuclei is due to a peculiar interaction of an external neutron pair with a target nucleus. The process (2) is possible in a peripheral interaction in which a break-up of external neutrons occurs without a secondary particle production. The determination of the process probability and measurement of the angular distribution of produced α particles is one of the major tasks of $^6$He nucleus experimental studies. Such events are presented in emulsion by an image of a track scattering of double charged particle

$$^6He \rightarrow {^4He} + 2n \qquad (2).$$

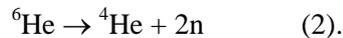

Analogous $^6$Li nucleus interactions in emulsion with α-particles in the final state were observed due to reactions $^6$Li → $^4$He + $^2$H and $^6$Li → $^4$He + $^1$H+n in ref. 4.

Table 2. Mean multiplicity values in events without projectile nucleus fragments.

| Nucleus | <Q> | <$n_\pi$> | <$n_h$> | Momentum (AGeV/c) |
|---|---|---|---|---|
| $^4$He | 1.2±0.1 | 2.3±0.1 | 4.7±0.2 | 4.5 |
| $^6$Li | 2.0±0.1 | 4.0±0.1 | 6.1±0.2 | 4.5 |
| $^6$He | 1.7±0.3 | 2.2±0.4 | 5.5±1.0 | 2.67 |

Indeed, in the present study, on a 457 cm scanned path of double charged tracks it was found 23 inelastic interactions having secondary particles (see for example fig. 5) and, in addition, five cases of a gradual change of the direction behind a scattering point. The mean momentum in these five events is equal to 15.6±3.8 GeV/c, and after scattering to 9.1±2.6 GeV/c. The scattering angle θ do not exceed 0.35° in all the found cases, and the mean transverse momentum of α-particles <$p^\alpha_T$> is about 0.035 GeV/c. We notice that in dissociation processes $^6$Li → $^4$He + $^2$H and $^6$Li → $^4$He + $^1$H+n at a $^6$Li nucleus momentum 4.5A GeV/c the corresponding <$p^\alpha_T$> is about 0.15 GeV/c [3,4]. A narrower transverse momentum distribution in the process (2) points to on a very peripheral picture of α-particle production in a coherent interaction.

A mean free range defined on the 457 cm path taking into account the registered coherent interactions is λ($^6$He)=16.3±3.1 cm. This value is larger than the corresponding one for a $^6$Li nucleus determined in papers [2-5] as 14.3± 0.3 cm. It might be assumed that the excessive value of λ($^6$He) can be explained by an assumption of a 50% identification efficiency of the process (2). The validity of such an assumption will indicate that the contribution of coherent interactions to the total cross-section exceeds 20%.

Table 3. Mean transverse momenta of projectile double charged fragments produced in incoherent interactions

| Projectile nucleus | <$p^\alpha_T$> (MeV/c) | <$p^{3He}_T$> (MeV/c) | Momentum (AGeV/c) |
|---|---|---|---|
| $^{12}$C | 238±8 | | 4.5 |
| $^4$He | 239±12 | 223±20 | 4.5 |
| $^6$Li | 144±10 | 136±10 | 4.5 |
| $^6$He | 90±15 | 180±20 | 2.67 |

The mean transverse momentum values of <$p^\alpha_T$> derived from an emulsion analysis are presented in Table 3. A significantly smaller value for the $^6$He nucleus case might be considered as an indication that in these events interactions of external nucleons proceed on a larger distances than the corresponding interactions with a $^6$Li nucleus. The table contains the transverse momenta of $^3$He nucleus fragments. Comparable values of <$p^\alpha_T$> for $^6$He and $^6$Li nuclei point to a similar interaction mechanism of α particle cores for these nuclei.

Summarizing this part, we conclude that the registration of nucleus-nucleus interaction vertex allows one to select a process of the coherent neutron chopping. At the same time, we notice that the study of angular distributions in an angular range less than 0.5° is complicated. To select such a reaction in emulsion one needs to perform momentum measurements both in a particle entrance path and in an exit one which increases dramatically the work with a measuring microscope.

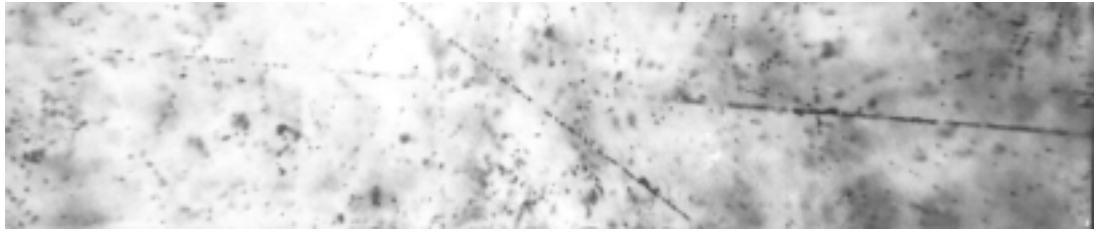

Fig. 6. Event of charge exchange $^3$H→$^3$He without charged meson production.

### Interactions of $^3$H nuclei and charge exchange process of $^3$H nuclei into $^3$He nuclei in emulsion

It has already been mentioned that the emulsion stack has been irradiated in a nucleus beam with $^3$H nucleus dominating contribution. There is an opportunity to extent our study for this nucleus. At present, it has been found 66 inelastic interactions containing secondary charged particles on a scanned 1565 cm track path. The mean free path corresponding to these data is equal to 23.7±3.0 cm. It should be noticed that invisible neutron chopping events are not taken into account.

Charge exchange cases of $^3$H nucleus into $^3$He one have been found on this path. Such a process is easily distinguishable in emulsion due to a visual ionization density increasing from a single charged track to a double charged one. Multiple scattering measurements of primary single charged tracks and subsequent double charged ones have confirmed approximate momentum conservation.

For CE studies searches for double charge tracks on a middle and exit part of an emulsion stack have been performed. The found tracks were followed to their entrance in emulsion or to the production point. In this way, 100 % searching efficiency was confirmed for $^6$He nucleus tracks. Besides, 36 double charged particle tracks produced in interactions of $^3$H nuclei with emulsion nuclei have been found.

A CE process without charged meson production (3) has been observed in 22 events among them (see for example fig. 6)

$$^3H + A\,(Z) \to {^3He} + A\,(Z-1)\,(+\pi^o) \qquad (3).$$

There are single charged particle tracks in the remaining events and these events belong to CE process (4) with charged meson production

$$^3H + A\,(Z) \to {^3He} + \pi^- + A\,(Z-1) \qquad (4).$$

The mean free path for the CE process $^3$H→ $^3$He is $\lambda(^3H+Em\to {^3He})=40\pm16$ m corresponding to the effective cross-section of a process in emulsion 3.1±1.2 mb. The number of beam $^3$H nuclei is equal to $(1.8\pm0.4)\cdot10^4$. It is defined via the number of found $^6$He nucleus tracks and the ratio of $^3$H nuclei to $^6$He ones. The emulsion thickness passed by a $^3$H nucleus is accepted to be 9 cm.

Table 4 presents the mean transverse momenta of $^3$He nuclei in various interaction classes characterized by the absence or presence of a single charged particle $n_\pi$ and the mean h-particle multiplicity $<n_h>$. In all events registered $n_\pi$ does not exceeds unity while the number of h particles does not exceed 4. Notice that the overwhelming majority of h particles are slow particles with ranges in emulsion less than 3 mm.

Table 4. Mean transverse momenta of $^3$He nuclei for various interaction classes

| Interaction $n_\pi$ - $<n_h>$ | Number of events | % | $<p_t>$, GeV/c | Reference |
|---|---|---|---|---|
| 0 - 0 | 9 | 25 | 0.15±0.03 | the present paper |
|  | 673 | 40 | 0.19±0.06 | [6] |
| 1 – 0 | 6 | 17 | 0.18±0.04 | the present paper |
|  | 568 | 34 | 0.37±0.06 | [6] |
| 0 - 1.85 | 13 | 36 | 0.27±0.05 | the present paper |
| 0 - 1.25 | 272 | 16 | 0.36±0.08 | [6] |
| 1 - 1.25 | 8 | 22 | 0.23±0.04 | the present paper |
| 1 - 1.25 | 164 | 10 | 0.54±0.07 | [6] |

Table 4 shows that charge exchange processes with and without charged meson production have approximately an equal probability. In spite of small event statistics it is possible to conclude that the number of events with target nucleus excitation exceeds the one without excitation. In the first two event groups the values $<p_t>$ are significantly less than in the subsequent ones with target nucleus excitation. Probably, the momentum difference in the first two groups indicates that in the first one ($n_\pi=0$ and $n_h=0$) there is a contribution to CE process without meson production in a CE process with neutral meson production, and such a process occurs with reduced momentum transfers.

Table 4 also shows the results of paper [6] on the study of a charge exchange process of $^3$H nucleus to $^3$He one on a nucleus Mg target at a momentum 3A GeV/c performed on the GIBS spectrometer with particle charge sign determination. Events without a negative particle are referred to the class $n_\pi=0$ and those without a positive particle to $n_\pi=1$. Mean positive particle multiplicities are shown in the column $<n_h>$. The event ratio with and without charged meson production obtained in this paper is in agreement with the result of the paper [6]. The event fraction with target nucleus excitation obtained in this paper significantly exceeds the corresponding value for events with positive particles in GIBS spectrometer data [6]. The major reason for this may be explained by an incomplete registration of slow fragments in the spectrometer with an extended target. One might consider authors' notice of ref. [6] as a confirmation of such a supposition that recorded positive particles appear to be protons with a mean energy 80 MeV. This circumstance points to their non evaporation generation mechanism while in emulsion a major fraction of heavily ionizing particles are slowly interacting particles with a range less than 3 mm (protons within such a range have an energy less than 26 MeV). Indeed, the agreement with GIBS' data is reached if in the emulsion events only events having range over 3 mm (the so-called *g* particles) instead of all heavily ionizing *h* particles. The obtained mean values of $^3$He the nucleus transverse momenta are less than in GIBS data in all the discussed interaction groups. This shows that an additional verification of registration efficiency of $^3$He nuclei at large emission angles is needed.

Concluding this section, we would like to notice good prospects for studies of the CE process $^3$H→ $^3$He in emulsion. Such an experiment appears to be the simplest one from the technical viewpoint and the analysis of the appropriate data requires considerably less time than in case of other experiments in emulsion. In this case, search for interactions is performed by backward scanning of only registered $^3$He nucleus tracks up to an interaction vertex rather than by following all the beam nucleus tracks entering emulsion. This method considerably reduces the time of accumulation of the interaction statistics.

**Conclusion**

1. A secondary beam containing $^6$He nuclei with momentum 2.67A GeV/c has been produced on an external beam of the JINR Synchrophasotron by charge exchange of $^6$Li nuclei. The beam profile has been formed to provide uniform emulsion irradiation. Effective use of emulsion increased by a factor of 3 – 4 due to enlarged working area with a uniform track density. The optimal profile beam has allowed one to increase effective use of emulsions and provide better conditions of work with a microscope. Such an experience ought to be applied in the future.
2. A test irradiation of an emulsion stack is performed in a beam of relativistic $^6$He nuclei. The results of determination of $^6$He nucleus momenta via measurements of a multiple Coulomb scattering has pointed to a sufficiently reliable identification of relativistic $^6$He nuclei in emulsion. This makes it possible to study processes like $^6$Li and $^6$He nuclei charge exchange, $^6$He nuclei production in fragmentation of heavier projectiles.
3. Information on the topology of inelastic interactions of relativistic $^6$He nuclei in emulsion has been obtained. Events of $^6$He inelastic interactions having $\alpha$ particles in the final states and without any other secondary charged particles are registered. Narrow distributions of $\alpha$ particles in a coherent production process may be due to the circumstance that external neutron interactions occur in very peripheral events with target nuclei. We notice that experiments with registration of a nucleus-nucleus interaction vertex allow one to select a process of coherent chopping of neutrons from $^6$He nuclei.
4. A process of charge exchange of $^3$H and $^3$He has been registered. The mean free path for such a process in emulsion is equal to $\lambda(^3\text{H}+\text{Em}\to {}^3\text{He})=40\pm16$ m and the corresponding value of the effective cross-section in emulsion is 3.1±1.2 mb. The charge exchange process has approximately the same cross-section with and without production of a charged meson. The mean transverse momentum of $^3$He nuclei in events without target nucleus excitation is equal to 0.16±0.03 GeV/c, while the one with excitation is 0.24±0.05 GeV/c. The mean free path of $^3$He nuclei for inelastic interactions in emulsion is equal to 23.7±3.0 cm.

The authors express their warmest thanks to Prof. A. I. Malakhov for his interest in this work and practical support. Valuable contributions to these results has been given by the Synchrophasotron operation personnel, LHE beam transport group, emulsion chemical proceeding group, technical staff responsible for scanning and measurements in emulsion. We would like to mention the contribution of A. V. Pisetskaya (FIAN, Moscow) to a microscope visual analysis. We are also indebted to N. M. Piskunov (LHE JINR), F. G. Lepekhin, and B. B. Simonov (PNPI, Gatchina) for fruitful discussions.